\documentstyle[11pt,epsf]{article}
\input psfig
\def\beq{\begin{eqnarray}}
\def\eeq{\end{eqnarray}}
\def\bea{\begin{eqnarray*}}
\def\eea{\end{eqnarray*}}

\def\centeron#1#2{{\setbox0=\hbox{#1}\setbox1=\hbox{#2}\ifdim
\wd1>\wd0\kern.5\wd1\kern-.5\wd0\fi
\copy0\kern-.5\wd0\kern-.5\wd1\copy1\ifdim\wd0>\wd1
\kern.5\wd0\kern-.5\wd1\fi}}
\def\ltap{\;\centeron{\raise.35ex\hbox{$<$}}{\lower.65ex\hbox{$\sim$}}\;}
\def\gtap{\;\centeron{\raise.35ex\hbox{$>$}}{\lower.65ex\hbox{$\sim$}}\;}

\def\singleandthirdspaced{\baselineskip=\normalbaselineskip\multiply
    \baselineskip by 130\divide\baselineskip by 100}

\newcommand{\newc}{\newcommand}
\newc{\qbar}{{\overline q}}
\newc{\Kahler}{K\"ahler }
\newc{\deltaGS}{\delta_{\rm GS}}
\begin{document}
\begin{titlepage}
\begin{flushright}
{\large hep-th/yymmnnn \\ SCIPP-2006/13\\
}
\end{flushright}

\vskip 1.2cm

\begin{center}

{\LARGE\bf Gauge Mediation in Metastable Vacua}

\vskip 1.4cm

{\large  Michael Dine and John Mason}
\\
\vskip 0.4cm
%{\it $^a$Stanford Linear Accelerator Center,
%     Stanford CA 94309} \\
{\it Santa Cruz Institute for Particle Physics and
\\ Department of Physics,
     Santa Cruz CA 95064  } \\
%{\it $^c$Physics Department,
%     University of California,
%     Santa Cruz CA 95064  }

\vskip 4pt

\vskip 1.5cm

\begin{abstract}
Until recently, dynamical supersymmetry breaking seemed an exceptional
phenomenon, involving chiral gauge theories with a special structure.  Recently
it has become clear that requiring only metastable states with broken supersymmetry
leads to a far broader class of theories.  In this paper, we extend these
constructions still further, finding new classes which, unlike earlier theories,
do not have unbroken, approximate $R$ symmetries.  This allows construction of
new models with low energy gauge mediation.

\end{abstract}

\end{center}

\vskip 1.0 cm

\end{titlepage}
\setcounter{footnote}{0} \setcounter{page}{2}
\setcounter{section}{0} \setcounter{subsection}{0}
\setcounter{subsubsection}{0}

%%%%%%%%%%%%%%%%%%%%%%%%%%%%%%%%%%%%%%%%%%%
%%%%%%%%%%%%%%%%%%%%%%%%%%%%
\singleandthirdspaced

\section{Introduction:  Retrofitting O'Raifeartaigh Models}

Until recently, dynamical supersymmetry breaking seemed an exceptional phenomenon
\cite{dsb}.
Analysis of the Witten index indicated that such breaking can only occur in chiral
gauge theories, and even then only under rather special circumstances.
Recently, however, Intriligator, Shih and Seiberg\cite{iss} exhibited a class of vector-like
gauge theories which possess metastable, supersymmetry-breaking minima.
Feng,
Silverstein, and one of the present authors showed that this phenomenon is quite
common\cite{dfs}.  One can take a generic Fayet-Iliopoulos model, and simply replace
the scales appearing there with dynamical scales associated with some underlying,
supersymmetry-conserving, dynamics.

Consider, for example, a theory with chiral fields, $A$, $Y$ and $Z$, and superpotential:
\beq
W = \lambda Z (A^2 - \mu^2) + m YA.
\label{simplemodel}
\eeq
This is a theory which breaks supersymmetry.  The scale, $\mu$, can be generated dynamically
by introducing a dynamical gauge theory and replacing $\mu^2$ by the expectation value
of some suitable composite operator.  One simple possibility is to take the extra sector
to be a pure gauge theory, say $SU(N)$, and introduce a coupling
\beq
\int d^2 \theta {Z \over 4 M} W_\alpha^2.
\label{zgauge}
\eeq
This structure can be enforced by discrete symmetries.  The gauge theory has a
$Z_N$ discrete symmetry, so if $A$ and $Y$ transform like $W_\alpha$,
while $Z$ is neutral, the only couplings of dimension three or less which are invariant
are those above.
Integrating out the gauge fields, leaves a superpotential:
\beq
W = \lambda Z A^2 + {\Lambda^3 e^{-8 \pi Z/b_0} \over M} + mYA
\eeq
The resulting potential has a minimum at $Z \rightarrow \infty$, i.e. it exhibits
runaway behavior.  But the Coleman-Weinberg corrections give rise
to a local minimum at $Z=0$.

This simple theory can be used in an interesting way
as a hidden sector in a supergravity theory.  Previously, most known models
of dynamical supersymmetry breaking contained no gauge singlets.  As a result,
one could not write dimension five operators giving rise to gaugino masses,
and the leading contributions arose from anomaly mediation.\footnote{An exception
was provided by the Intriligator-Thomas models\cite{it,yanagida}.  In general, naturalness
arguments would require an anomalous discrete R symmetry in these theories, but
this will be required in our models below, as well.  Models with local minima
with broken supersymmetry have been considered in the past as well, e.g.
\cite{dimopoulos,luty}}
But in these ``retrofitted" models,
there is no obstruction to the existence of a coupling of $Z$
to the various gauge fields, so there is no difficulty generating gaugino masses.
One still faces the problem of large potential flavor violation.

Various strategies were discussed in \cite{dfs} to break supersymmetry at lower scales.
However, it was difficulty within the examples presented there, to build
realistic models.  (Alternative strategies
based on the ISS models were put forward in \cite{banks}.)
The difficulty is illustrated by our simple model.
At low energies, the model has a continuous $U(1)$ R symmetry.  Under
this symmetry, the fields $Y$ and $Z$ carry charge two.
  This symmetry forbids
gaugino masses.  Within simple models with chiral fields only, it is difficult
to find examples where the analogs of the $Y$ and $Z$ fields acquire expectation
values, breaking the symmetry. 

On the other hand, at least since Witten's work on the ``inverted hierarchy" long
ago\cite{inverted},
gauge interactions have been
known to destabilize the origin of moduli potentials in O'Raifeartaigh models.  In
this note, we exhibit examples with gauge symmetries where the potential has a local
minimum away from, but not far away from, the origin.  The symmetry is broken
and a rich phenomenology is possible.

In the next section, we generalize further the constructions of \cite{dfs}.  It is
troubling that the simplest ``retrofitted" models introduce additional mass
parameters in the lagrangian, and we explain how these can also be understood
dynamically, in terms of a single set of gauge interactions.
In section three, we introduce the gauged model,
compute the potential, and verify that $R$ symmetry is broken for a range of
parameters.  We then turn to the construct of low energy (direct) models of gauge
mediation.  We implement a solution of the $\mu$ problem which following on
ideas of Giudice, Rattazzi and Slavich\cite{grmu,giudicenew}.  In the conclusions we discuss issues of
stability, fine tuning, and directions for future work.

\section{Naturalness in the Retrofitted Models}

The model of eqn. \ref{simplemodel} has, in addition to $\mu^2$, the dimensionful parameter $m$.
Some strategies to obtain both mass terms
dynamically were discussed in \cite{dfs}, but these often
run afoul of naturalness criteria.  A simple variant of the ideas above works here as well,
however.  Suppose, again, that one has a pure gauge theory with a large scale, $\Lambda$.
Then the scales $\mu^2$ and $M$ can be replaced by couplings:
\beq
W_\Lambda = Z A^2+{1 \over M_p^4} {Z W_\alpha^4} + {1 \over M_p^2} W_\alpha^2 A Y.
\eeq
We have taken the scale here to be the Planck scale, but
one could well imagine that some other large scale (the GUT scale, for example)
would determine the size of these operators.
Now
\beq
\mu^2 = {\Lambda^6 \over M_p^4};  M = {\Lambda^3 \over M_p^2}
\eeq
and $M$ and $\mu$ are naturally of the same order.  This structure can readily be compatible
with a discrete $Z_N$ R-symmetry.  For example, if $\alpha$ is an $N$'th root of unity,
\beq
W_\alpha^2 \rightarrow \alpha W_\alpha^2; Z \rightarrow \alpha^{-1} Z; 
A \rightarrow \alpha A; B\rightarrow \alpha^{-1} B.
\eeq
It is also necessary to impose a $Z_2$, under which $A$ and $B$ are odd, to prohibit
the coupling $A^2 B$ (additional restrictions may be necessary for particular values of $N$).
In the model we consider in the next section, the extra $Z_2$ is not necessary; the
gauge symmetries forbid the unwanted coupling.

In \cite{dfs}, still another mechanism to obtain dimensional parameters naturally was
described:  it was shown that one can naturally obtain Fayet-Iliopoulos terms.
We will not exploit this in our model building in this paper, but this may also be
a useful tool.

\section{Including Gauge Interactions:  Coleman-Weinberg Calculation}

The basic model is a $U(1)$ gauge theory, with charged fields $Z^{\pm}$
and $\phi^{\pm}$, and a neutral field, $Z^0$.  The superpotential
of the model is 
\beq
W= M_+ Z^+ \phi^- + M_- Z^- \phi^+ + \lambda Z^0(\phi^+ \phi^- -\mu^2).
\eeq
The model breaks supersymmetry.  For simplicity, we take
$M_+ = M_- =M$.  If $\vert M^2 \vert < \vert \lambda^2 \mu^2 \vert$,
at the minimum of the potential:
\beq
\phi^+ = \phi^- = v~~~~~~v^2 = {\lambda^2 \mu^2 - M^2 \over \lambda^2}
\eeq
(up to phases) while
\beq
F_{Z^+} = F_{Z^-} = M v;~~F_{Z^0} = {M^2 \over \lambda}.
\eeq
There is a flat direction with
\beq
Z^{\pm} = -{\lambda Z^0 \phi^{\pm} \over M}.
\eeq
As in the previous section, both the parameters $M$ and $\mu$ can arise from
dynamics at some much larger scale; the structure can be enforced by discrete
symmetries.

It is easy to compute the potential at large $Z$.  In this limit, the theory is
approximately supersymmetric, and the gauge fields, as well as certain linear combinations of
the $Z$'s, are massive.  It is then possible to integrate out the massive fields,
writing a {\it supersymmetric} effective action for the light fields.  
It is also helpful to work in a limit of large $M$, $M \gg \lambda \mu$,
so that the $F$ components of $Z^{\pm}$ are larger than that of $Z^0$.
Any supersymmetry breaking should show up from the $F$ components of the various 
fields, i.e. as terms of the form
\beq
\int d^4 \theta (Z^{+ \dagger} Z^+ f+ Z^{-\dagger} Z^-g + Z^{0\dagger} Z^0 h),
\eeq
where $f,g,h$ are functions of the $Z$'s.

So we need to compute Feynman diagrams (supergraphs) with external $Z$ fields.  These
are greatly simplified by isolating
the pieces of the form $F^\dagger F$,
noting that the external lines all have two $\theta$'s.
We need the propagators in a generalized 't Hooft-Feynman gauge:
\beq
<Z^+ Z^{+\dagger}> = {i \over p^2 -M_V^2} \times ~(1 + {\rm \theta-dependent})
~~~~<V(\theta_1) V(\theta_2)> = -{1 \over 2}{i\delta(\theta_1-\theta_2) \over
p^2 - M_V^2}.
\eeq
There are essentially two graphs.  In the diagram with two external $Z^0$'s
an internal $\phi^+$ and $\phi^-$ (with mass $\lambda Z^0$), the theta's at the
interaction vertices are soaked up by the $F$'s, and the graph is simply
\beq
\lambda^2 \vert F_{Z^0}\vert^2\int {d^4 p \over (2 \pi)^4}{1 \over (p^2 - 
\vert \lambda^2 Z^{0}\vert^2)^2}
= {\lambda^2 \over 16 \pi^2} \vert F_{Z^0}\vert^2 \ln(\vert \lambda Z^0 \vert^2).
\eeq
For the gauge interactions, the diagrams are equally simple.  The leading
interaction is $Z^{\pm \dagger} Z^{\pm} (2gV)$.  Now one has $\int d^4 \theta$
at each vertex, but the delta function in the gauge boson propagator soaks up the
remaining $\theta$'s).  So we have
\beq
-2 g^2 (\vert F_{Z^+} \vert^2 + \vert F_{Z^-} \vert^2) \int {d^4 p \over (2 \pi)^4p^2(p^2-M_V^2)}
\eeq
$$=~~~-{4 g^2 \vert F_{Z^+} \vert^2 \over 16 \pi^2}  \ln(g^2 \vert Z^+\vert^2).$$
So overall, the asymptotic behavior of the potential is given by:
\beq
V = {1 \over 16 \pi^2}\left (\vert F_{Z^0}\vert^2 \lambda^2 - 4 g^2
\vert F_{Z^+} \vert^2 \right )\ln(Z^{02}/\Lambda^2)
\eeq
for a cutoff, $\Lambda$.

\begin{figure}[htbp]
\centering 
\centerline{\psfig{file=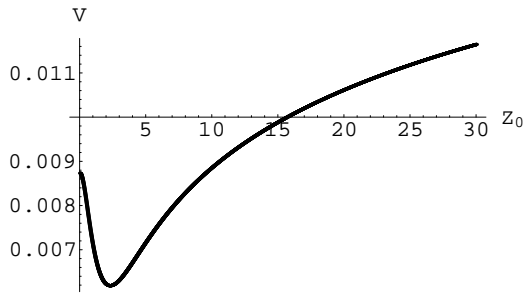,angle=-0,width=7cm} }
\caption{Z potential with $g=.4,\lambda=1,M=1,\mu=1.5$.}
\label{zpotential}
\end{figure}

For a range of $g$ and $\lambda$, then, the potential grows at large $Z^0$.
We wish to determine whether, within this range, there is a range for which
the potential has negative curvature at small $Z$.  The answer is yes.
At small $Z$, it is simplest to do the Coleman-Weinberg calculation directly.
In fig. \ref{zpotential}, we have plotted the potential for several values of $g$ and $\lambda$,
and, indeed, for a range of parameters, there is a minimum at non-zero $Z^0$.  

\begin{figure}
\begin{center}
$\begin{array}{cc} 
\epsfxsize=2in
\epsffile{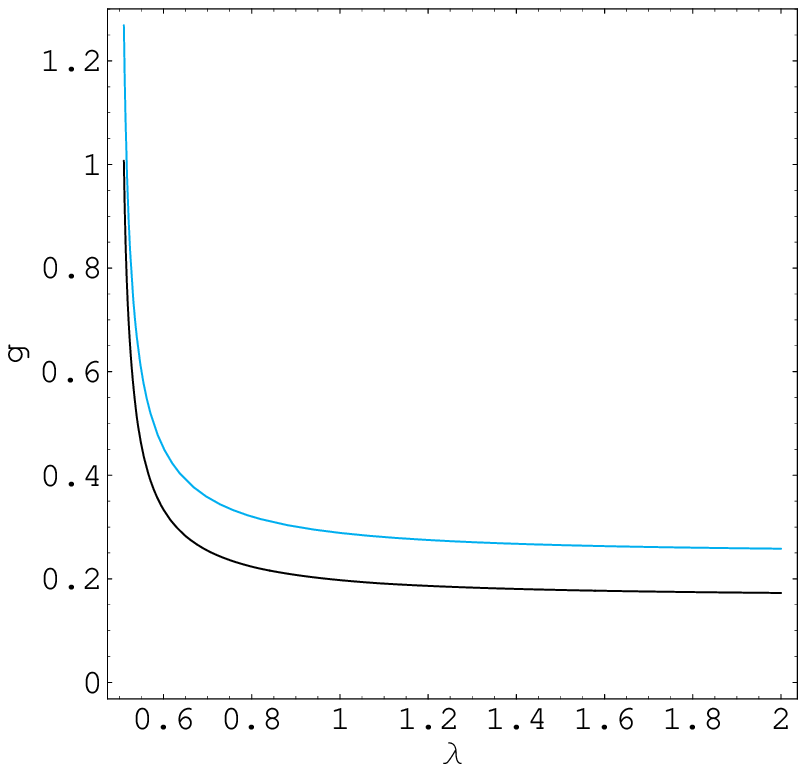} &
\epsfxsize=2in
\epsffile{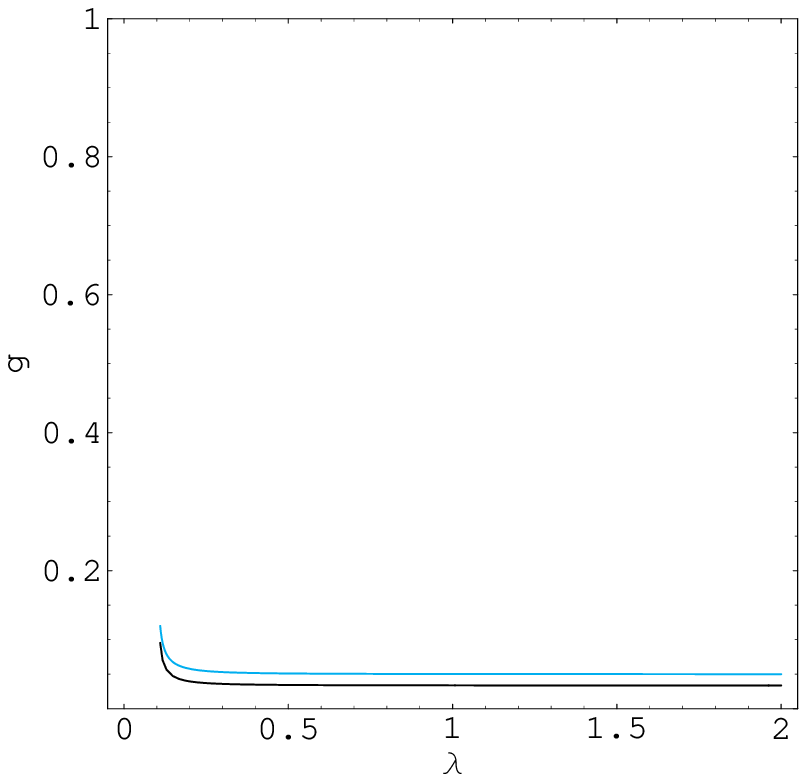} \\
(a) & (b) \\
\epsfxsize=2in
\epsffile{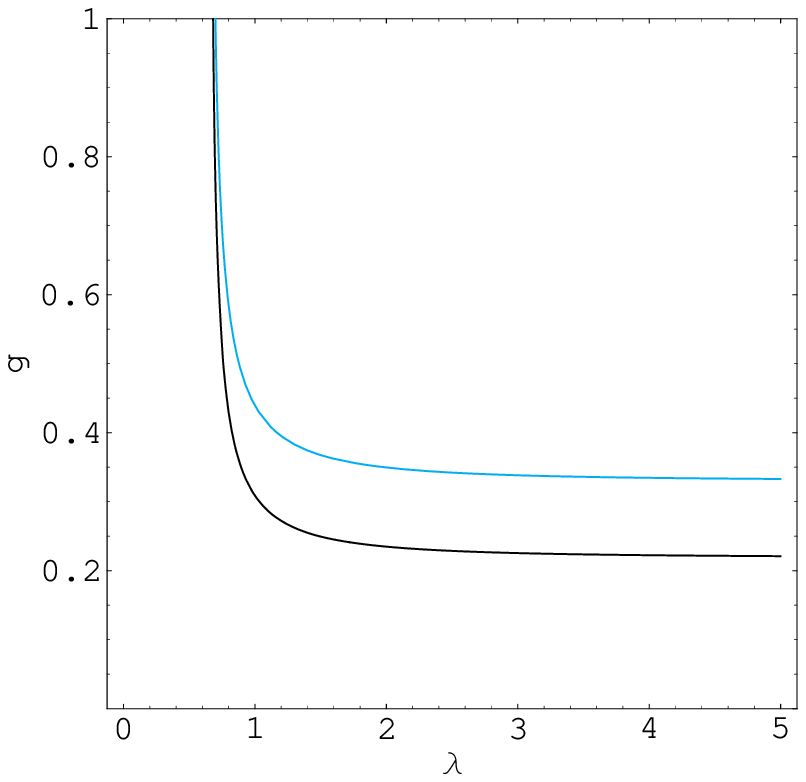} &
\epsfxsize=2in
\epsffile{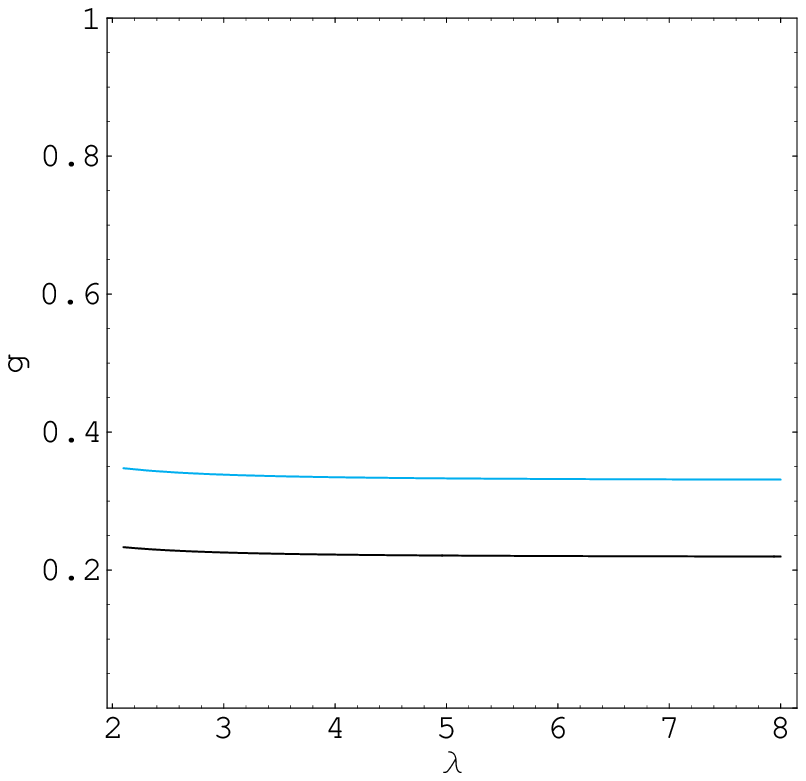} \\
(c) & (d) \\
\end{array}$
\end{center}
\caption{Graphs of the regions of parameter space where there is a local
R-breaking minimum. Below the blue line is the region
where the potential grows positive at large $Z_0$ and above the black line
is the region where the potential curves down at the origin.
The values of $h$ are $h=.5$, $h=.1$, $h=.66$  and $h=2$ in (a),(b),(c), and (d),
respectively}
\label{tuningfigure}
\end{figure}

In general, for small $Z^0$, 
\beq
V(Z) = {\rm const} + m_Z^2 \vert Z^0 \vert^2.
\eeq
The constant is obtained by diagonalizing the full mass matrix.  The condition that
$m_Z^2 < 0$ os a condition on the ratio of gauge to Yukawa couplings, as is the 
condition that the potential should rise at $\infty$.  The bands of allowed $g$ and $\lambda$,
for different values of $h$, are indicated in fig. \ref{tuningfigure}.

Note that $Z^0, Z^\pm$ at the minimum are of order $\mu$.  So by dialing the
dynamical scale, one can obtain supersymmetry breaking at any scale.
In addition, the accidental $R$ symmetry of the low energy theory
is spontaneously broken.  The would-be Goldstone boson
gains a substantial mass once couplings to supergravity are included, as
explained in \cite{baggerrandall}.  This, then, is just the sort of structure
we would like in gauge mediated models\cite{gaugemediation}.
We can now couple $Z^0$, for example,
to messenger fields.  We will initiate a study
of such models in the next section.

\section{Low Energy, Direct Mediation}

In the model we have described above, the scale of supersymmetry breaking is a free parameter,
and the scale of $R$-symmetry breaking is of the same order.  This is clearly only one
of a large class of possible models.  By making different choices of charges, for example,
one can avoid introducing the dimensionful parameter, $M$.  (One can, and in general should,
introduce two independent mass parameters in the original model).  One can now make
a model of direct gauge mediation by introducing a set of messenger fields, $M$ and $\bar M$,
with the quantum numbers of a $5$ and $\bar 5$ of $SU(5)$, and with couplings to $Z^0$:
\beq
\lambda^\prime Z^0 \bar M M.
\eeq
Then the standard gauge mediation computation
yields a positive mass-squared for squarks and sleptons.  Note that in this case, if $Z^0$
is neutral under the discrete $R$ symmetry, no symmetry forbids a large mass term
for $\bar M M$.  This problem arises because of our choice of coupling
in equation \ref{zgauge}.  In order that symmetries forbid a $\bar M M$
mass term, one needs that $Z$ transform non-trivially under the discrete
symmetry, as discussed in section 2.

One might worry that in this model, in addition to the far away supersymmetric minimum,
there is a close by one with 
\beq
\lambda^\prime \bar M M -\lambda \mu^2 =0.
\eeq
For suitable $\lambda$ and $\lambda^\prime$, our candidate minimum remains a local
minimum of the potential, however; it is also sufficiently metastable.  To see this,
note that in the metastable minimum, the quadratic terms in the $M, \bar M$ potential are:
\beq
\lambda^\prime F_{Z^0}^* \bar M M + ~{\rm c.c.}~ + \vert \lambda^\prime
Z^o\vert^2 (\vert M \vert^2
+ \vert \bar M \vert^2).
\eeq
We can work in a regime where $\lambda \mu \gg M$.  In this regime, at the 
minimum
\beq
Z^{\pm} \sim \mu~~~~Z^0 \sim M/\lambda ~~~~~F_{Z_0} \sim {M^2 \over \lambda}
\label{roughscaling}
\eeq
so the curvature of the $M,\bar M$ potential is positive provided $\lambda^\prime \gg \lambda$.  
(The first relation in eqn.
\ref{roughscaling} follows from the fact that the potential for $Z^\pm$ will exhibit structure
on the scale of the $\phi^\pm$ expectation values; the second from the vanishing
of $F_{\phi^\pm}$.)  Numerically, we find that the situation is better than this; $F_{Z^0}$
is typically significantly smaller than $\vert Z_0 \vert^2$, even when $\lambda \mu \sim M$.

Note that the energy difference between the metastable and the supersymmetric vacuum
is of order:
\beq
\Delta E = {M^2 v^2}.
\eeq
The barrier height, on the other hand, is of order the shift in $\phi^\pm$ times the $\phi^\pm$
masses in the metastable minimum, or
\beq
V_0 \sim \lambda^2 \mu^4.
\eeq
So a thin walled treatment is appropriate\cite{coleman},
and the bounce action is at least as large as
\beq
S \sim {\pi^2 S_1^4 \over 2 \Delta E^3}
\eeq
where
\beq
S_1 \sim {\lambda  \mu^3}.
\eeq
This gives an estimate for the bounce action:
\beq
S \sim \pi^2 \lambda^{-2} \left ({\lambda^6 \mu^6 \over M^6} \right ).
\eeq
$\lambda$ and $M^2 /\mu^2 \lambda^2$ were, by assumption, our small parameters,
and the decay amplitude can easily be extremely small, even if the small parameters
are not.

The problems of generating suitable $\mu$ and $B_\mu$ terms in
gauge-mediated models are well-known.
If we simply couple $Z^0$ to $\bar H H$, with a small coupling,
this can generate a small $B_\mu$ but this will lead to a very small $\mu$ term.
In this framework, we can also generate a $\mu$ and $B_\mu$ term of a reasonable
order of magnitude, following the ideas of
\cite{grmu} and \cite{giudicenew}.
This approach involves introducing a singlets, $S$ with couplings similar
to those of the
NMSSM\cite{grmu}.  It is also necessary to double the messenger sector, i.e. to have fields
$M_i, \bar M_i$, $i=1,2$.
We can take the additional terms in the superpotential to be (we avoid
giving names to all of the the various couplings at this stage):
\beq
W= Z^0(y_1 M_1 \bar M_1 + y_2 M_2 \bar M_2)+ h S M_1 \bar M_2
+ S^3.
\label{sw}
\eeq 
The renormalizable couplings can be restricted to this form
by a discrete $R$ symmetry.  For example, taking
\beq
\alpha= e^{2 \pi i \over N}
\eeq
and supposing $Z^0$ transforms with phase $\alpha^{-1}$, we can take:  
\beq
M_1 \rightarrow \alpha^2 M_1~~
\bar M_1 \rightarrow \bar M_1 ~~ M_2 \rightarrow
 M_2 ~~ \bar M_2 \rightarrow \alpha^2 \bar M_2
~~ S \rightarrow \alpha^{-3} S
%%%S^\prime \rightarrow \alpha^2 S^\prime
~~ H_U H_D \rightarrow \alpha^{-1} H_U H_D. 
\eeq
This still allows some dangerous couplings; in particular, $Z^0 S^\prime$.
This can be forbidden by an additional $Z_2$, for example, under which
$S^{\prime}$ is odd, and one of $H_U$ or $H_D$ is odd.

In \cite{grmu,giudicenew}, it was shown that the one loop corrections to the $S$ mass
vanish (to order $F^2$) in a model such as this, and the two loop contributions can be negative.
As a result, the $S$ vev is one loop order, the $\mu$ term one loop order,
and the $B_\mu$ term two loop order.

\section{Conclusions}

This paper can perhaps be viewed as the culmination of the program initiated by ISS\cite{iss}.
ISS exhibited vector-like models with non-vanishing Witten index, in which
there are metastable states in which supersymmetry is dynamically broken.  Ref.
\cite{dfs} enlarged the set of possible models, by simply taking O'Raifeartaigh theories
and replacing all mass parameters by dynamically generated scales. As noted in
\cite{dfs}, because
these models often contain singlets neutral under gauge symmetries
and discrete $R$ symmetries, they open
up new possibilities for building supergravity models with supersymmetry
broken dynamically in a hidden sector.  This may be particularly interesting
in light of recent studies of the {\it landscape} of string vacua\cite{landscape}.
In many string
constructions,  large chiral theories of the type previously thought needed for
dynamical supersymmetry breaking seem rare.  Non-chiral theories with
singlets seem far more common.   $R$-symmetries may be rare, however\cite{dinesun}.

One can ask about the cosmology of the susy-breaking
vacuum states.  First, there is the issue of metastability.
We would expect the susy breaking
states in the retrofitted models to
be highly metastable.  Even before coupling to
gravity, the supersymmetric minimum lies far away in field space, and the
amplitudes are suppressed by huge exponentials.
Before worrying about gravity, the decay amplitudes are typically extremely
small,
\beq
\Gamma \sim e^{-c M^4/\mu^4},
\eeq
where $c$ is a number of order 1.
Once coupled to gravity, the standard Coleman-DeLuccia
analysis will give vanishing amplitude in most cases for decays to big crunch spacetimes.
A number of papers have appeared recently discussing the question:  can the system
find its way into the metastable vacuum.  In the ISS case, if one assumes that the system
was in thermal equilibrium after inflation, one finds that the broken susy minimum,
is thermodynamically
favored\cite{thermal}.  In the O'Raifeartaigh models, the same can be true; the
analysis is in fact simpler.  For example, in our
models, we have large numbers of messenger fields, which are light in the metastable vacuum.
More generally, the low energy theory has accidental, approximate symmetries near the origin
of field space, and even non-thermal effects (e.g. in cosmologies with low reheating 
temperatures) may favor these points.

Our principle interest in this paper was to
construct models of direct mediation.  For this, in both
the models of \cite{iss} and \cite{dfs} there was an obstacle:
the low energy theories possessed an accidental continuous $R$ symmetry.  In this paper, we have
shown how, by adding gauge interactions, one can break the $R$ symmetry spontaneously.
It is straightforward to add messengers to implement dynamical supersymmetry breaking.
In this way, one could construct models with susy breaking scale as low as $10$'s
of TeV.  One still must confront the standard difficulties in gauge mediation, especially
the $\mu$ problem and the question of fine tuning.
We have considered one mechanism
for solving the $\mu$ problem in these theories, and argued that it is technically
natural.  The models, at low energies, look like the conventional NMSSM.  Ameliorating
the tuning problems will require a hidden sector with more fields, in which the
squarks are not parameterically heavy compared to the doublet sleptons.  This question
will be explored elsewhere\cite{mason}, as will further studies of the $\mu$ term.

\noindent
{\bf Acknowledgements:}
We thank Tom Banks, Jonathan Feng and Eva Silverstein for conversations.
We are grateful to R. Rattazzi, Nathan
Seiberg, David Shih and Yuri Shirman  for criticisms of an
early version of this paper.
This work supported in part by the U.S.
Department of Energy.  M. D. thanks the Kavli Institute for Theoretical
Physics for hospitality during the course of this project.

\end{document}